\documentclass[pra,showpacs,twocolumn,superscriptaddress]{revtex4}

\usepackage{amsfonts}
\usepackage{amstext}
\usepackage{amssymb,amsmath,graphicx,enumerate,color}
\usepackage{hyperref}

\newcommand{\de}[1]{\left( #1 \right)}
\newcommand{\De}[1]{\left[ #1 \right]}
\newcommand{\DE}[1]{\left\{ #1 \right\}}

\newcommand{\ket}[1]{\left| #1 \right\rangle}

\newcommand{\mean}[1]{\left\langle #1 \right\rangle}
\newcommand{\op}[1]{{\mathbf{#1}}}

\newcommand{\esp}[1]{{\mathsf{#1}}}

\newcommand{\tr}{\mathrm{Tr}}

\newcommand{\sig}{\op{\sigma}}
\newcommand{\s}{\op{S}}
\newcommand{\CCC}{{\mathbb{C}}^2\otimes{\mathbb{C}}^2\otimes{\mathbb{C}}^2}

\newcommand{\ie}{{\it{i.e.}}: }

\newcommand{\ajp}[4]{{\it{Am. J. Phys.}} {\bf{#2}}, #3 (#4)}

\newcommand{\jmp}[4]{#1 {\it{J. Math. Phys.}} {\bf{#2}}, #3 (#4)}
\newcommand{\jpa}[4]{{\it{J. Phys. A: Math. Gen.}} {\bf{#2}}, #3 (#4)}

\begin{document}

\title{Tomographic characterization of three-qubit pure states with
only two-qubit detectors}
\author{D. Cavalcanti}\email{dcs@fisica.ufmg.br}
\affiliation{Departamento de F\'{\i}sica, CP 702, Universidade Federal de Minas Gerais, 30123-970, Belo Horizonte, Brazil}
\author{L.M. Cioletti}\email{leandro@mat.ufmg.br}
\affiliation{Departamento de Matem\'atica, CP 702, Universidade Federal de Minas Gerais, 30123-970, Belo Horizonte, Brazil}
\author{M.O. \surname{Terra Cunha}}\email{tcunha@mat.ufmg.br}
\affiliation{Departamento de F\'{\i}sica, CP 702, Universidade Federal de Minas Gerais, 30123-970, Belo Horizonte, Brazil}
\affiliation{Departamento de Matem\'atica, CP 702, Universidade Federal de Minas Gerais, 30123-970, Belo Horizonte, Brazil}

\begin{abstract}
A tomographic process for three-qubit pure states using only
pairwise detections is presented. 
\pacs{03.67.-a,03.65.Wj,03.65.Ud}
\end{abstract}

\maketitle

The understanding of multipartite entanglement is one of the objectives of the quantum information community.
 Bipartite entanglement is well understood for pure states, where the Schmidt decomposition\cite{sch}
 plays the central role. Also much has been learnt about mixed states in recent years\cite{dagmar}. The problem one has to face when the numbers of
 parties grow is the consequent increasing in the complexity of the correlations involved.
Recently, a significant progress in three-qubit correlations has been reported by N. Linden, S. Popescu and W. K. Wootters\cite{LPW}, who have shown that ``{\emph{Almost every pure state of three qubits is completely determined by its two-particle  reduced density
matrices}}''. It is important to emphasize the context in which the
work is developed which is understanding how information is stored in multipartite systems. Specifically, they show that there is
no more information on the three-party state than what is already
contained in the three reduced pair states. 
Their result is, at first sight,
surprising. In the case of two qubits it is not difficult to show that
several global states give the same reduced states. For bipartite
pure states, the Schmidt decomposition theorem\cite{sch} asserts
that, given a vector $\ket{\Psi} \in \esp{V} \otimes \esp{W}$, one
can choose orthonormal basis $\DE{\ket{v_i}}$ for $\esp{V}$, and
$\DE{\ket{w_j}}$ for $\esp{W}$ such that $\ket{\Psi} = \sum _k
\lambda _k \ket{v_k}\otimes \ket{w_k}$. In the case of two qubits,
$\esp{V}$ and $\esp{W}$ have dimension $2$ and the vector state
can be written as\footnote{It is possible (and usual) to include the phase $e^{i\phi}$ on the
basis vectors, however we are exactly interested in its indeterminacy from the
local states, which justifies this unusual Schmidt decomposition.}
\begin{equation}
\ket{\Psi\de{\theta ,\varphi}} = \cos \theta \ket{v_1}\otimes \ket{w_1}
+ e^{i\varphi} \sin \theta \ket{v_2}\otimes \ket{w_2},
\label{Schm}
\end{equation}
with $\theta \in \De{0,\frac{\pi}{4}}$ and $\varphi \in
\De{0,2\pi}$. Writing the density matrix in the basis
$\DE{\ket{v_i}\otimes\ket{w_j}}$ makes it clear that the relative
phase $e^{i\varphi}$ is locally inaccessible, {\it{i.e.}}, for
fixed $\theta$ all reduced density matrices are equal. So, there
is more information on the full two-qubit state than on the
parts represented by the reduced states. As a clarifying example,
the two Bell states\cite{Bel} $\ket{\Phi _{\pm}} = \DE{\ket{00}
\pm \ket{11}}/\sqrt{2}$ originate the same local
states\footnote{In fact the other two Bell states $\ket{\Psi _{\pm}} =
\DE{\ket{01} \pm \ket{10}}/\sqrt{2}$ also generate the same local states as
$\ket{\Phi _{\pm}}$, but for another reason: degeneracy in Schmidt
decomposition corresponding to $\theta = \frac{\pi}{4}$ in eq.\ (\ref{Schm}).
}.

Suppose now that an experimental physicist wants to make
tomographic measurements of three qubits with only two
detectors. Reference
\cite{LPW} shows that all necessary information is available in the two-qubit reduced matrices,
 but does not suggest any practical method to do it.
In this work we present a tomographic protocol for the complete characterization of
generic three-qubit pure states, based only on pairwise detections.
The protocol works whenever the result of Ref. \cite{LPW} holds,
 \ie for all pure states except for a restricted class that will be comment latter (see eq.~(\ref{GHZfamily})).
  It must be said that although coincidence measurements are allowed,
  only local operations should be done, \ie one can think of three distinct laboratories,
   two of them equiped with detectors at each time, and a classical electronic coincidence line between them. In this sense, we are only implementing LOCC (Local Operations with Classical Comunnication).


First of all, let us review the authors argument. Consider an
arbitrary pure state $\ket{\nu}=\sum _{ijk} \nu _{ijk} \ket{ijk}$ of
three qubits A, B, and C. A general state (pure or mixed) that has
the same reduced states of $\ket{\nu}$ can be obtained through a
pure state $\ket{\Psi}$, describing three qubits plus an
environment $E$ (this process is called a {\emph{purification}},
and its existence can be shown by the Schmidt decomposition). The
fact that $\ket{\Psi}$ has the same reduced states than
$\ket{\nu}$   when restricted to each two-qubit subspaces puts
restrictions in its form. It is shown that for a generic state
$\ket{\nu}$, these restrictions determine the form of $\ket{\Psi}$ as
$\ket{\Psi}=\sum _{ijk} \nu _{ijk} \ket{ijk}\otimes\ket{E}$, where
the environment is factorized. Consequently the environment state is pure, and the three-qubit state
is necessarily $\ket{\nu}$. It is a very elegant argument, which
also allowed Linden and Wootters to generalize this results
for $N$ qubits\cite{LW}. It is, however, rather abstract, and gives
no clue for the experimentalist to completely characterize his
three-qubit pure state.

The route we will take uses the generalization of the well known
expression for one qubit:
\begin{equation}
\op{\rho} = \frac{1}{2} \sum_{\mu=0}^3 b_{\mu}\sig _{\mu},
\label{Bloch}
\end{equation}
where $\sig _0$ is the $2\times 2$ identity  matrix, and $\sig _i$ are the Pauli matrices
\begin{eqnarray}
\sig _0 = \left[
\begin{array}{cc}
1 & \\
& 1
\end{array} \right],
&\quad &
\sig _1 = \left[
\begin{array}{cc}
& 1 \\
1 &
\end{array} \right],
\nonumber \\
\sig _2 = \left[
\begin{array}{cc}
& -i \\
i &
\end{array} \right],
& \quad &
\sig _3 = \left[
\begin{array}{cc}
1 & \\
& -1
\end{array} \right],
\label{PauliM}
\end{eqnarray}
where we leave blank all null entries.
The coefficients can be obtained from the expression
\begin{equation}
b_{\nu} = \tr \DE{\op{\rho}\sig_{\nu}}.
\label{tomo1}
\end{equation}
It is important to observe that eq.\ (\ref{tomo1}) implies $b_0 = 1$
 (normalization of $\op{\rho}$)
 and that the complete characterization of the
 state can be achieved with three mean value measurements
  $b_{\nu} = \mean{\op{\sig _{\nu}}}$ (\ie by the three components of
  the so called Bloch vector). In fact, this is a tomographic scheme
  for determining a spin-$\frac{1}{2}$ state\cite{Wei9?}.

To generalize eq.\
(\ref{tomo1}) for three qubits, let us define
\begin{equation}
\s _{\gamma \mu \nu } = \sig _{\gamma} \otimes \sig _{\mu} \otimes
\sig _{\nu},
\end{equation}
and denote the state of three qubits by
\begin{equation}
\op{\rho} =\de{\frac{1}{2}}^3 a_{\gamma \mu \nu}\s _{\gamma \mu \nu},
\label{gen2}
\end{equation}
where we have adopted the convention of summation over repeated indexes throghout the paper
(latin indexes from $1$ to $3$;
 greek indexes from $0$ to $3$). Once again, the coefficients
$a_{\gamma \mu \nu}$ can be obtained tomographically by the
relation
\begin{equation}
a_{\gamma \mu \nu} = \tr \DE{\op{\rho}\s _{\gamma \mu \nu}}.
\label{tomo2}
\end{equation}
A first important consequence of eq.\ (\ref{tomo2}) is that $a_{000} =
1$. As Pauli matrices are traceless, the reduced density operators
are given by
\begin{subequations}
\label{rhopairs}
\begin{eqnarray}
\op{\rho}_{BC} =& {\tr} _A (\op{\rho}_{ABC}) =& \frac{1}{4}a_{0 \mu \nu}\s _{\mu
\nu},\\
\op{\rho}_{AC} =& {\tr} _B (\op{\rho}_{ABC}) =& \frac{1}{4}a_{\gamma 0
\nu}\s _{\gamma \nu},\\
\op{\rho}_{AB} =& {\tr} _C (\op{\rho}_{ABC}) =&
\frac{1}{4}a_{\gamma \mu  0}\s _{\gamma \mu},
\end{eqnarray}
\end{subequations}
where $\s _{\mu\nu}=\sig _{\mu} \otimes \sig _{\nu}$.

To directly determine $\op{\rho}$ through eq.~(\ref{tomo2}) one needs to evaluate sixty three mean values.
Nine of them $\de{3\times 3}$ are the three components of each Bloch vector ($a_{i00}$, $a_{0j0}$, $a_{00k}$),
 and can be determined by individual detections.
  Twenty seven $\de{3\times 9}$ are the pair correlations ($a_{ij0}$, $a_{i0k}$, $a_{0jk}$)
  and must be obtained through two-qubit coincidence measurements.
  The remaining twenty seven are three-qubit correlations,
   and are directly available only through three-qubit coincidence detections.
   For a general mixed state the number of mean values to be determined is exactly the same
    as the number of coefficients in the density operator.
However, any previous knowledge on the state of the system is expected to reduce the number of
parameters needed. In particular, for pure states we can use the idempotency relation,
\begin{equation}
\op{\rho}^2 = \op{\rho},
\label{idempot}
\end{equation}
to obtain the coefficients $a_{ijk}$ from those available in the pair states.
From expression (\ref{idempot}) we get 64 equations,
which can be organized in four sets: the first set consists of one solely equation
\begin{subequations}
\label{tomoeqs}
\begin{equation}
\sum_{ijk} (a_{i00}^2+a_{0j0}^2+a_{00k}^2+a_{ij0}^2+a_{i0k}^2+a_{0jk}^2+a_{ijk}^2)=7,
\label{eqsum}
\end{equation}
the second set is given by
\begin{equation}
3a_{i00} = a_{ij0}a_{0j0}+a_{i0k}a_{00k}+a_{ijk}a_{0jk},
\label{eq2o}
\end{equation}
with similar equations under permutations of indexes.
\begin{widetext}
The third set is consituted of
\begin{equation}
3a_{ij0}=a_{i00}a_{0j0}+a_{00k}a_{ijk}+a_{0jk}a_{i0k}
-\frac{1}{2}\epsilon_{ilt}\epsilon_{jmu}a_{lm0}a_{tu0}-\frac{1}{2}\epsilon_{il
t}\epsilon_{jmu}a_{tuk}a_{lmk},
\label{eq1o}
\end{equation}
together with analogous equations under cyclic permutations, and where
we use the Levi-Civitta symbol $\epsilon _{ijk}$ for the totally
antisymmetric tensor. Finally, the fourth group
\begin{equation}
3a_{ijk}=a_{i00}a_{0jk}+a_{0j0}a_{i0k}+a_{00k}a_{ij0}
-\epsilon_{ilt}\epsilon_{jmu}a_{tu0}a_{lmk}-\epsilon_{ilt}\epsilon_{knv}a_{t0v
}a_{ljn}-\epsilon_{jmu}\epsilon_{knv}a_{0uv}a_{imn}.
\label{matrix}
\end{equation}
\end{widetext}
\end{subequations}

The tomographic process is thus constituted by the thirty six mean values measured in individual and two-qubit coincidences, and the sixty four equations (\ref{tomoeqs}), that must be solved for $a_{ijk}$. The Linden, Popescu and Wootters' result guarantee the
generic solution of the  whole set of equations.

 Anyhow, the last set (\ref{matrix}) gives $27$ linear equations
  on the $27$ unknowns $a_{ijk}$.
  In case they are linearly independent, this specific set can give the complete solution. We  numerically checked such independence in all of hundreds of random choices. However, as pointed out by the authors of Ref. \cite{LPW}, there are exceptions, for states like
\begin{equation}
\ket{GHZ\de{\theta ,\varphi}} = \cos \theta \ket{000} + e^{i\varphi}\sin \theta \ket{111},
\label{GHZfamily}
\end{equation}
with $\theta \in \de{0,\frac{\pi}{2}}$,
in which the phase $e^{i\varphi}$ is pairwise inaccessible,
in the same sense as its analog in the two-qubit states (\ref{Schm}).
 Thus, all the exceptions are the states that, for some choice of local
 basis, can be written as (\ref{GHZfamily}), since for any other state,
  all the phases can be obtained without involving triorthogonal basis
  vectors. In such case, eqs. (\ref{matrix}) can not be linearly independent, but we conjecture that a more geometrical argument can show the generic independence, and also point out the exceptions (\ref{GHZfamily}).

A parameter counting shows how rare the exceptions
are. A vector in $\CCC$ is given by $8$ complex numbers (\ie $16$
real numbers). Normalization and global phase reduce it to $14$
real numbers. Local unitary operations are given by the action of
$SU\de{2}$ group in each qubit{\footnote{$SU\de{2}$ is used instead of
$U\de{2}$ because we have already eliminated the global phase.}}. $SU\de{2}$ is
parametrized by $3$ real numbers (the $3$ Euler angles, or the $3$ components of a
vector $\vec{v}$ for the Lie algebraic parametrization $U\de{\vec{v}} = \exp
\DE{i\vec{v}\cdot\vec{\op{\sigma}}}$), so the orbit space has real dimension $5$
(\ie $14 - 3\times 3$)\cite{LPS}. As the fase $\varphi$ in (\ref{GHZfamily}) can be changed by local unitarities, the GHZ family has just one parameter, $\theta$. So it is like a regular curve in a five dimensional manifold.

The protocol here presented can be described in the following steps: first two-qubit measurements determine  the $36$ mean values characterizing all the two-qubit reduced states (\ie the coefficients $a_{i00}$, $a_{0j0}$, $a_{00k}$, $a_{ij0}$, $a_{i0k}$, and $a_{0jk}$); then this experimental data is used as input on the $27$ linear equations (\ref{matrix}), and their solution generically determines the $27$ remaining coefficients $a_{ijk}$, \ie  generic three-qubit pure states are completely characterized by these $27$ calculated coefficients plus the $36$ directly measured ones. In fact, we can check for the purity of the measured state using the obtained coefficients to test the remaining $37$ equations  (\ref{eqsum}, \ref{eq2o}, \ref{eq1o}). If any of these ``testing'' equations is not satisfied (within experimental precision), one should conclude that the original state is mixed, and can not be determined with only $36$ mean values.

Some other questions can be raised on this issue. Is there any other tomographic process, restricted to two-qubit detections, that can determine the state with fewer measurements, without introducing new exceptions? Is the optimal number of measurements, $14$, achievable with this kind of restriction?
Recently Di\'osi \cite{Diosi} pointed out
that a generic tripartite pure state can be determined by the knowledge of any two constituent
pairs. Again this is a  {\emph{generic}} result in which interesting exceptions arise.
For example, for three qubits, if an experimentalist decides to directly access  $\rho_{AB}$ and
$\rho_{BC}$, and the prepared state is
$\ket{\psi} =  \frac{1}{\sqrt{3}}(\ket{000} + \ket{010} + e^{i\varphi}\ket{111})$, it will
be impossible to determine the phase $\varphi$. It would be available, however, at  $\rho_{AC}$.
In fact, it is an interesting problem  to classify the multipartite pure states by the partial information
necessary to completely determine them. Such a classification could help understanding
the curious geometric structure behind pure states. 

In this paper we provide for a feasible experimental prescription to completely characterize a
generic three-qubit pure state using only two-qubit detectors.

\begin{acknowledgments}
DC and LMC are supported by the Brazilian agency CNPq.
MOTC thanks Paulo Nussenzveig for the hospitality at USP, where part of the work was done.
The authors thank  Marcelo Fran\c ca  Santos for stimullating discussion and Maria Carolina Nemes for comments on a previous version of this article.
\end{acknowledgments}


\begin{thebibliography}{0}
\bibitem{sch}E.~Schmidt, {\it{Math. Ann.}} {\bf{63}}, 433 (1907); A.~Ekert and P.~L.~Knight, \ajp{}{63}{415}{1995}; P.~K.~Aravind, \ajp{}{64}{1143}{1996}.

\bibitem{dagmar}C.~H.~Bennett, D.~P.~DiVincenzo, J.~A.~Smolin, and W.~K.~Wootters, \pra {\bf{54}}, {3824}{(1996)}; D. Bru\ss, \jmp{}{43}{4237}{2002}.

\bibitem{LPW}N.~Linden, S.~Popescu, and W.~K.~Wootters, \prl {\bf{89}}, {207901}{(2002)}.


\bibitem{Bel}J.~S.~Bell, {\emph{Speakable and unspeakable in quantum mechanics}}, (Cambridge Univesity Press, Cambridge, 1987).


\bibitem{LW}N.~Linden and W.~K.~Wootters, \prl {\bf{89}}, {277906}{(2002)}.

\bibitem{Wei9?}K.~Vogel and H.~Risken, \pra {\bf{40}}, {R2847}{(1989)}; S.~Weigert, \pra {\bf{45}}, {7688}{(1992)}; J.~-P.~Amiet and S.~Weigert, \jpa{``Reconstructing the density matrix of a spin $s$ through Stern-Gerlach measurements,''}{31}{L543}{1998}.


\bibitem{LPS}N.~Linden, S.~Popescu, and A.~Sudbery, \prl {\bf{83}}, {243}{(1999)}.

\bibitem{Diosi}L.~Di\'osi, \pra {\bf{70}}, {010302(R)}{(2004)}.


\end{thebibliography}
\end{document}